%% file: Renyi.v3.tex
\documentclass[prb,aps,letterpaper,10pt,twocolumn,showpacs]{revtex4-1}
\usepackage{graphicx}
\usepackage[final,dvips]{epsfig}
\usepackage{color}
\usepackage{dcolumn}  
\usepackage{bm}       
\usepackage{array}    
\usepackage{amssymb,amsmath}
\usepackage{appendix}
\usepackage{booktabs}
\usepackage[colorlinks,citecolor=red]{hyperref}
\include{MyCommand}

\usepackage{dcolumn}  
\newcolumntype{d}[1]{D{,}{,}{#1}}
\newcolumntype{.}[1]{D{.}{.}{#1}}
\usepackage{natbib}

\def \Renyi {R\'{e}nyi }  

\begin{document}

\title{Entanglement properties of the antiferromagnetic-singlet transition 
in the Hubbard model on bilayer square lattices}

\author{Chia-Chen Chang, Rajiv R. P. Singh, and Richard T. Scalettar}

\affiliation{Department of Physics, University of California Davis, CA 95616, USA}

\begin{abstract}
We calculate the bipartite \Renyi entanglement entropy of an $L\times L\times 2$ bilayer Hubbard model
using a determinantal quantum Monte Carlo method recently proposed by Grover [Phys. Rev. Lett. {\bf 111}, 
130402 (2013)]. 
Two types of bipartition are studied: (i) One that divides the lattice into two $L \times L$ planes,
and (ii) One that divides the lattice into two equal-size ($L\times L/2\times 2$) bilayers. We compare 
our calculations with those for the tight-binding model studied by the correlation matrix method. As 
expected, the entropy for bipartition (i) scales as $L^2$, while the latter scales with $L$ with possible 
logarithmic corrections. The onset of the antiferromagnet to singlet transition shows up by a saturation 
of the former to a maximal value and the latter to a small value in the singlet phase. We comment on 
the large uncertainties in the numerical results with increasing $U$, which would have to be overcome
before the critical behavior and logarithmic corrections can be quantified.
\end{abstract}


\maketitle

\section{Introduction}
There are several reasons for recent excitement in the condensed matter 
theory community regarding quantum entanglement entropies of many-body 
lattice models.\cite{cardy,rmp_review} These entropies connect widely disparate 
fields of physics such as quantum information theory, quantum gravity and 
black holes with the many-body description of quantum phases and material 
science. When a macroscopic system is divided into two parts, the bipartite 
entanglement entropies provide universal signatures of quantum phase transitions 
and critical phenomena\cite{cardy,rmp_review,max} and have also been used to
demonstrate existence of topological quantum phases in spin 
models.\cite{levin_wen,kitaev,isakov,kagome,Jiang2012,Depenbrock2012} 

In a typical eigenstate of a many-body Hamiltonian, these entropies are 
extensive i.e., in a large system they are proportional to the volume of 
the system and are closely related to the thermal entropy.\cite{deutsch13,santos,michelle}
However, in the many-body ground state, such entropies typically obey 
an ``area-law'',\cite{hastings} that is they scale with the measure of the 
``area'' or boundary between subsystems. Such an ``area-law'' provides the 
basis for novel approaches for quantum many-body systems including 
density matrix renormalization group (DMRG) and their higher-dimensional 
tensor-network generalizations.\cite{White1992,Schollwock2005,tensor-network}

Despite this progress, unbiased numerical calculations of entanglement entropies 
in interacting lattice models of spatial dimensionality greater than one remain
a big challenge. Over the last few years several methods have been developed for 
quantum spin models, which go beyond the very small finite-systems for which exact 
eigenstates, reduced density matrices and entanglement entropies can be calculated 
by standard Lanczos type methods. The first such method is the quantum Monte Carlo
(QMC) method, which allows unbiased stochastic simulation of large 
systems.\cite{melko-swap,Humeniuk2012}
A second method is that of series expansions, where entanglement entropies are 
obtained as a power-series expansion in a suitable coupling 
constant.\cite{Singh2011,Kallin2011,oitmaa} 
A third method is that of numerical linked cluster expansion (NLCE), where the 
entanglement properties of the thermodynamic system are expressed as a sum over the 
contributions from different sized clusters, which can be evaluated numerically 
through exact diagonalization.\cite{nlc,nlc-dmrg} 

Different methods have their advantages and disadvantages. The QMC method deals 
with finite systems and its convergence can be rigorously established by sufficient 
sampling. One then needs an extrapolation to the thermodynamic limit. Since rather 
large system sizes can be simulated many quantities can be calculated with high 
accuracy. The series expansion method is particularly suitable for a system in 
which a small parameter exists. In that limit, it provides highly accurate answers 
in the thermodynamic limit. However, critical points necessarily lie at the boundary 
of the convergence radius of the series, and thus studying critical properties requires 
the use of series extrapolation methods.\cite{book} Both QMC and series expansion
methods are suitable for calculating \Renyi entropies of low integer order. In contrast 
the NLCE method, can be used to calculate any index \Renyi or von Neumann entanglement 
entropy. Series expansions and NLCE are also particularly useful for studying entanglement 
contributions from corners and other subleading manifolds as those contributions can 
be analytically isolated from those of other larger boundaries.

Unbiased calculations of entanglement properties of interacting lattice fermion 
systems in dimensionality greater than one, have only recently been initiated. 
In a system with only bilinear fermion terms in the Hamiltonian, the correlation matrix method
provides a very efficient method.\cite{Eisler2007,Peschel2009,Song2012} 
This technique is very powerful, allowing for calculations of ground state, excited state, 
finite-temperature or non-equilibrium entropy. All \Renyi or von Neumann entropies can 
be computed with similar ease. Recently, an unbiased approach for treating interacting
fermion systems was proposed by Grover.\cite{Grover2013} It uses the determinant 
quantum Monte Carlo (DQMC) method to calculate the low integer \Renyi 
entropies.\cite{Grover2013,Assaad2013} This is the method we employ here.

The bilayer Hubbard model is a particularly simple model that is known to have several 
phase transitions.\cite{Bouadim2008,Kancharla2007,Ruger2014} In the large $U$ limit, it reduces 
to the bilayer Heisenberg model, which has been extensively studied by quantum Monte Carlo 
simulations and other methods.\cite{heisenberg-qmc,Weihong1997,Hamer2012} When the interlayer 
coupling is weak the model has a N\'eel ordered phase. When the interlayer coupling is strong, it 
has a spin-gapped singlet phase. 
The two phases are separated by a second phase order transition. It has been shown numerically 
that the transition is in the university class of the three-dimensional classical Heisenberg 
model.\cite{heisenberg-qmc}

The entanglement entropy offers a potentially very interesting way of studying this model.
In addition to the possible signature of the phase transition, the entanglement entropy
can also provide a measure of the Fermi surface properties of the system. It is well known
that in non-interacting fermion systems, there is a logarithmic breakdown of the ``area-law''.
Furthermore, this breakdown can be related via the Widom conjecture to quantitative 
geometrical features of the Fermi-surface relative to the boundary partitioning the two 
subsystems.\cite{klich,wolf,haas,swingle,sidel,michelle} Thus, entanglement entropy can
provide a direct evidence for a Fermi surface and hence a metallic phase in the model.

\begin{figure}
\includegraphics[scale=0.32]{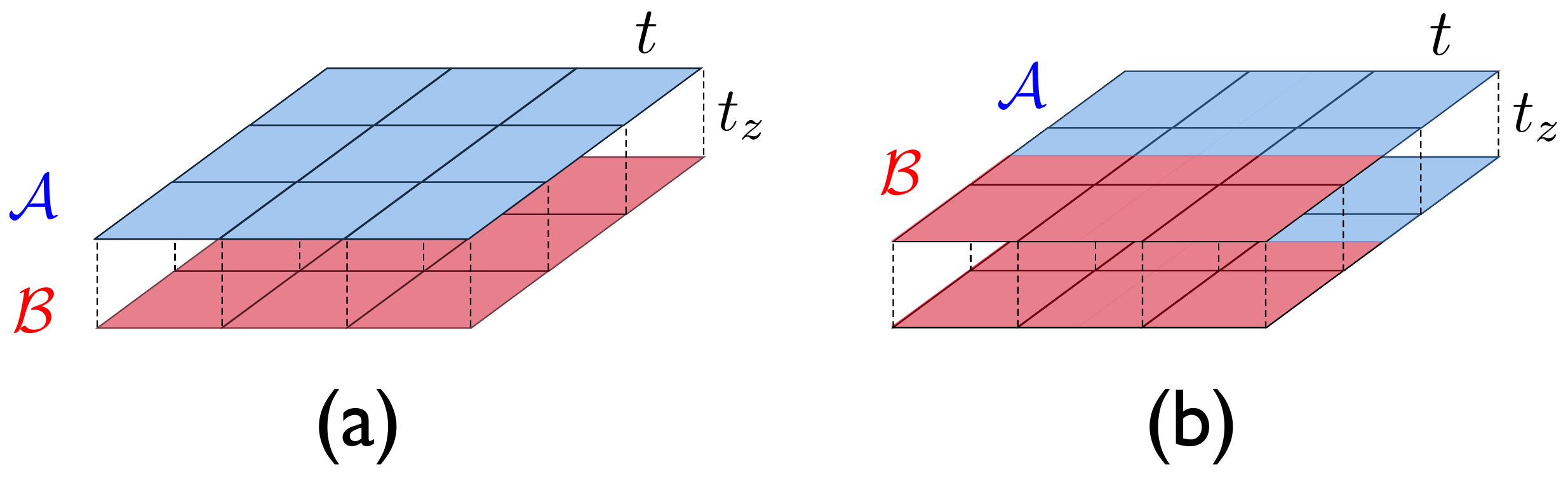}
\caption{(Color online)
Subsystem partitions studied in this work: (a) single layer, (b) two half-layers.
The corresponding entanglement entropies are denoted by $S_2^s$ and $S_2^h$ respectively.}
\label{fig:partitions}
\end{figure}

In this work, we study entanglement properties in the antiferromagnet (AF) to band insulator 
(BI) transition of the bilayer Hubbard model. We study two bipartitions of the lattice, where the 
lattice is divided into two planes and the other where the system is split up into two halves 
along one of the axis of the square-lattice, as illustrated in Fig.~\ref{fig:partitions}.
Our study of the tight-binding model further confirms the Widom conjecture. 
One can also see the metal to band-insulator transition and associated singularity in 
the tight-binding model study. For the Hubbard model, entanglement properties are examined 
as a function of interlayer hopping. We also discuss challenges in extracting critical 
properties of the \Renyi entropy and suggest possible solutions.

\begin{figure}
\includegraphics[scale=0.37]{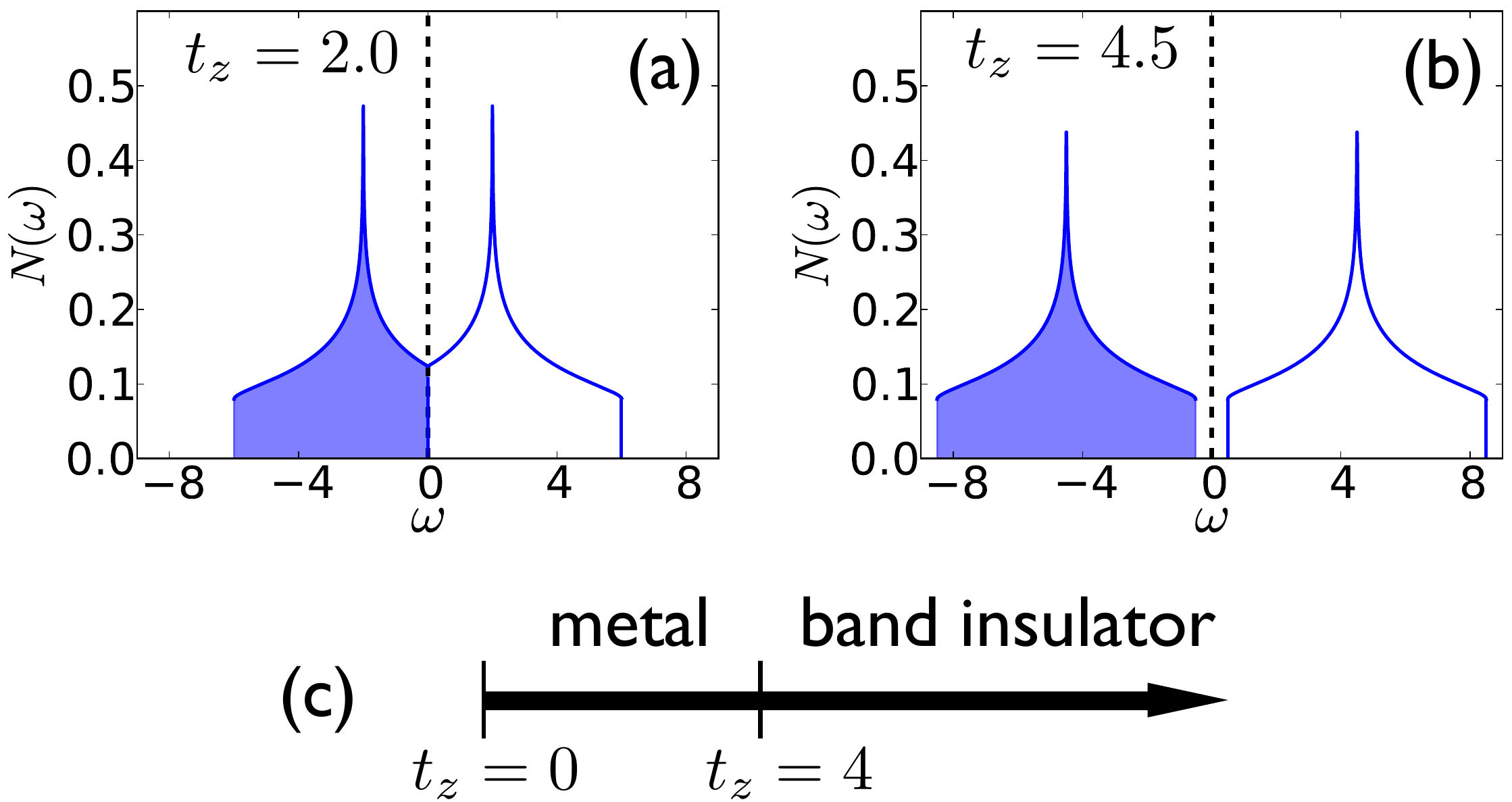}
\caption{(Color online) Panel (a) and (b): Non-interacting density of states of the half-filled 
bilayer system at $t_z=2.0$ and $t_z=4.5$ rspectively. The Fermi level is 
at $\omega=0$. (c): Non-interacting phase diagram of the model 
Eq.~(\ref{eq:Ham}).}
\label{fig:dos}
\end{figure}

\section{Bilayer square lattice Hubbard model}

The bilayer square lattice Hubbard model is defined by the following Hamiltonian
\begin{align}
  H =& -t \sum_{\ell\sigma}\sum_{\ob{\bfi\bfj}}\,
         \left( \ca{\bfi\ell}{\sigma} \de{\bfj\ell}{\sigma} + h.c. \right) \nonumber\\
     & -t_z \sum_{\bfi\sigma} \left( \ca{\bfi1}{\sigma} \de{\bfi2}{\sigma} + h.c. \right) \nonumber\\
     & + U \sum_{\bfi\ell} \left(n_{\bfi\ell\upa} - \frac 1 2 \right) 
                           \left(n_{\bfi\ell\dna} - \frac 1 2 \right) 
      - \mu\sum_{\bfi\ell\sigma} n_{\bfi\ell\sigma}.
     \label{eq:Ham}
\end{align}
Here $\ca{\bfi\ell}{\sigma}$ ($\de{\bfi\ell}{\sigma}$) creates (annihilates) an electron 
at site $\bfi$ with spin $\sigma\in\{\upa,\dna\}$ on an $L\times L\times 2$ lattice.
$\ell\in\{1,2\}$ is the layer index. 
$t$ and $t_z$ are intra- and inter-layer hoppings respectively. $U>0$ is the onsite repulsion, 
and the chemical potential $\mu$ determines the density of the system. We measure the energy 
in units of $t=1$. The chemical potential is kept at $\mu=0$ 
so that the system stays half-filled.

In the tight-binding limit, $U=0$, the physics of the system is determined solely 
by the inter-layer hopping $t_z$. As demonstrated in Fig.~\ref{fig:dos}, at $t_z \leq 4$ 
the system is in metallic phase with fully nested bonding and anti-bonding 
Fermi surfaces and finite density of states at the Fermi level. For $t_z > 4$, a gap opens up at the Fermi 
level and the system becomes a band insulator. 
The phase transition at $U=0$ is associated with the closing of the gap in the particle-hole
excitation spectra and this gap closes continuously.

At finite $U$, the model has been studied by several groups using numerical methods
such as QMC,\cite{Bouadim2008,Golor2014} dynamical mean-field theory (DMFT)\cite{Kancharla2007} 
and variational Monte Carlo (VMC).\cite{Ruger2014} These studies generally agree that
at large $U$, there is a direct transition from a singlet to a N\'eel phase as the
inter-layer hopping is varied.
However, properties of the model at small $U$ remain controversial. Is there a direct 
transition from a singlet to a N\'eel phase as inter-layer hopping matrix elements is 
varied? Both DQMC and DMFT studies suggest a paramagnetic metal phase. However, in roughly
the same parameter range, the VMC study predicts a N\'eel phase.
We will not address details of the phase diagram in this paper. 
Rather, we will examine entanglement properties of the system across the phase 
transition at small $U$ where there are no exact results.

\section{\Renyi entanglement entropy}

For a quantum many-body system divided into two disjoint subsystems $\cal A$ and $\cal B$,
one can defined a reduced density matrix for subsystem $\cal A$ by tracing out the degrees
of freedom in $\cal B$: $\rho_{\cal A} = \Tr_{\cal B} \left( \ket{\Psi}\bra{\Psi} \right)$,
where $\ket{\Psi}$ is the ground state of the total system. Then the \Renyi entanglement
entropy can be calculated from $\rho_{\cal A}$ as
\be
  S_n = \frac{1}{1-n}\log \left[ \Tr\, (\rho_{\cal A}^n) \right],
  \label{eq:Renyi}
\ee
where the von Neumann entropy can be recovered in the $n\rightarrow 1$ limit.
In this paper, we focus on the second \Renyi entanglement entropy, i.e. $n=2$.
For the bilayer square lattice, we consider two different subsystem partitions 
shown in Fig.~\ref{fig:partitions} and label the corresponding second \Renyi entropy 
as $S_2^s$ and $S_2^h$ respectively.

When there is no interaction, the model can be treated as two independent collections 
of spinless free fermions. 
In this case, the reduced density matrix $\rho_{\cal A}$ factorizes and, for each collection, 
the second \Renyi entropy can be expressed in terms of eigenvalues of the correlation 
matrix defined as\cite{Eisler2007,Peschel2009,Song2012}
\be
  C_{\bfi\bfj} = \ob{c_\bfi^\dagger c_\bfj^{\phantom{\dagger}}} 
  \label{eq:CorrMatrix}
\ee
where $\bfi,\bfj \in {\cal A}$, and $\ob{\ldots}$ denotes the expectation value with 
respect to the ground state. We numerically diagonalize the tight-binding Hamiltonian 
and construct $C_{\bfi\bfj}$ using the ground state orbitals.
Let $\lambda_k$ denote the eigenvalues of $C_{\bfi\bfj}$, then the \Renyi entropy $S_2$ 
is given by
\be
  S_2 = - \sum_k \log\left[\lambda_k^2 + (1-\lambda_k)^2 \right].
\ee

For interacting itinerant fermions, the correlation matrix method is not applicable and
measuring entanglement properties of the system often requires the knowledge of the ground
state wave function. Recently, there have been proposals of computing the \Renyi entropy
for lattice fermions using QMC technique that does not require the access of the ground 
state wave function. For example, by writing $S_2$ as a ratio of partition
functions, a generic scheme based on path integral Monte Carlo was developed in 
Ref.~\onlinecite{Humeniuk2012}. This method has been successfully applied to the single-band
Hubbard model in one dimension\cite{Bonnes2013}, and becomes the basis of a more recent
proposal of computing $S_2$ within the DQMC scheme.\cite{Broecker2014}

In this work, we adopt the scheme proposed in Ref.~\onlinecite{Grover2013}.
This technique exploits the fact that DQMC maps interacting fermions into a system of
free fermions coupled to fluctuating auxiliary fields. For a given set of auxiliary fields, 
the trace in Eq.~(\ref{eq:Renyi}) can then be carried out explicitly, and the second \Renyi 
entropy is expressed as\cite{Grover2013}
\begin{align}
  S_2 = & -\log \left\{
                \sum_{\{s\},\{s'\}}\, {\cal P}_s {\cal P}_{s'}\,
                \det \left[ \vphantom{\sum}\, G_{\cal A}(s) G_{\cal A}(s') \right.\right. \nonumber\\
        &            \qquad\left. \vphantom{\sum_{\{s\},\{s'\}}} \left.\vphantom{\sum}
                      + (\bfI - G_{\cal A}(s) )(\bfI - G_{\cal A}(s') )
                     \,\right]
              \right\}.
    \label{eq:S2Grover}
\end{align}
Here $\{s\}$ and $\{s'\}$ represent two sets of auxiliary fields. ${\cal P}_s$, ${\cal P}_{s'}$ 
are the probability distribution used to sample the fields. $G_{\cal A}(s)$, and similarly 
$G_{\cal A}(s')$, are one-particle Green's function whose spatial indices are restricted within
the subregion $\cal A$. $\bfI$ is the identity matrix. 
While the technique allows the measurement of higher order \Renyi entropies 
$S_n$,\cite{Grover2013,Assaad2013} it cannot directly access the von Neumann entanglement 
entropy. Measuring $S_n$ requires $n$ replica of subsystems. As a result,
the computation of $S_n$ is significantly more demanding, and not much more informative\cite{nlc}.
Therefore, in this work we only focus on $S_2$.

\section{Results and Discussions}

\begin{figure}[h]
\includegraphics[scale=0.43]{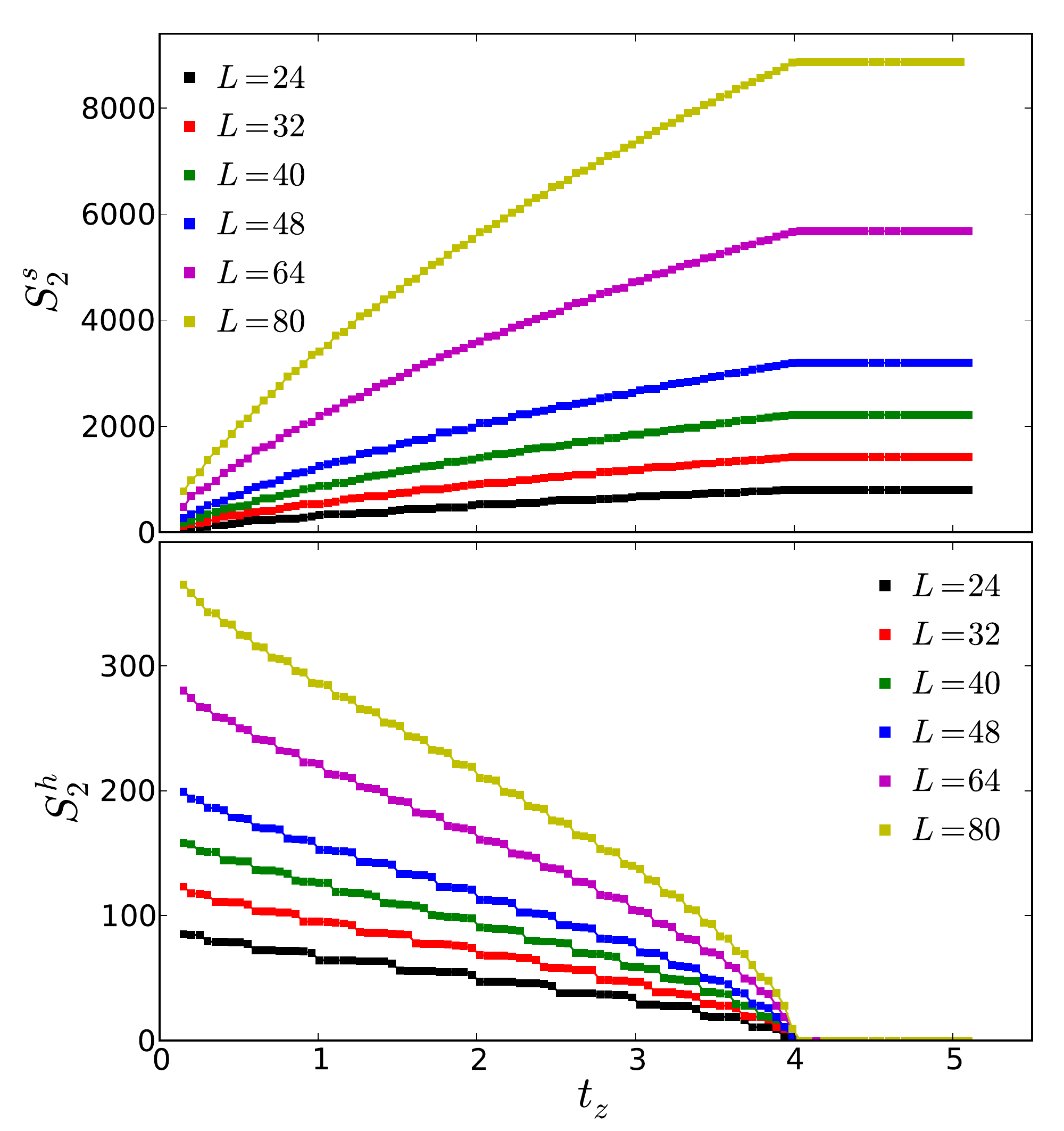}
\caption{(Color online)
The second \Renyi entropy as a function of inter-layer hopping $t_z$
for non-interacting electrons on bilayer square lattices at half-filling.
$L$ is the linear dimension of the bilayer lattice. The plateau structure
seen in the figures for $t_z < 4$ is caused by the ground state degeneracy 
at the Fermi surfaces of free electrons on finite lattices. 
}
\label{fig:S2.vs.tz.free}
\end{figure}

\subsection{Non-interacting bilayer model}

We first discuss entanglement properties for free electrons on the bilayer
square lattice. Fig.~\ref{fig:S2.vs.tz.free} summarizes the second \Renyi entropy 
as a function of $t_z$. Both $S_2^s$ and $S_2^h$ show a sharp signal at the 
critical point $t_z=4.0$. 
For the first partition where subsystem ${\cal A}$ is a single layer (c.f. Fig.~1), 
$S_2^s$ approaches the value $4(L-1) \ln 2$ when $t_z\rightarrow 0$, which, 
surprisingly, is different from $t_z=0$ case, where it vanishes identically.
To understand the results, let us recall that
for $t_z=0$, the ground state can be chosen independently in the two planes.
There should be no entanglement. However,
ground state degeneracy leads to finite entanglement for infinitesimal $t_z$.
To see this, consider the Fermi surfaces with $t_z$ set to zero. In each
plane, and for each spin component, the Fermi surface consists of the diamond-shape
boundary of the antiferromagnetic Brillouin zone at half-filling. 
For an $L\times L$ square lattice,
exactly $2L-2$ $\bfk$-points lie on the Fermi surface for each spin component.
Half of them will be occupied and the other half will be unoccupied.
This degeneracy is lifted by an infinitesimal $t_z$, which leads to equal number of
bonding and anti-bonding states with all the bonding states having negative energy 
and all the anti-bonding states having positive energy. Thus, at non-zero $t_z$, 
all the bonding states will be occupied and anti-bonding states empty.
Each occupied bonding state contributes $\ln 2$ to the entanglement entropy between 
the planes. There are $(2L-2)$ $\bfk$-points where such bonding states happen at 
the Fermi surface for each spin component. This gives us a total entanglement entropy 
between the planes of $4(L-1) \ln 2$.

In the insulating phase $t_z \geq 4.0$, the system consists of localized singlet pairs 
across the layers. We observe that $S_2^s=2L^2\ln 2$.
This is because for free electrons, each spin component (up or down) contributes 
$\ln 2$ to the entropy at each site since the bonding state is occupied. Thus, the
total entanglement entropy for the free fermion bilayer is $2L^2\ln 2$.
Alternatively, the free-fermion singlet state on every pair of sites of the bilayer
has $\ln 4$ entropy because the reduced density matrix has four equally likely 
choices: empty, spin-up, spin-down, both present. As a result, the total entropy
is also $L^2\ln 4 = 2L^2\ln 2$ as there are $L^2$ pairs of sites.

The second \Renyi entropy behaves quite differently for the second partition 
where $\cal A$ has two half-layers. In particular, $S_2^h$ decreases monotonically 
with increasing $t_z$. In the insulating phase where the lattice is filled with 
localized singlet pairs, $S_2^h=0$ since $\cal A$ and the rest of the lattice 
are decoupled completely.

\begin{figure}[h]
\includegraphics[scale=0.42]{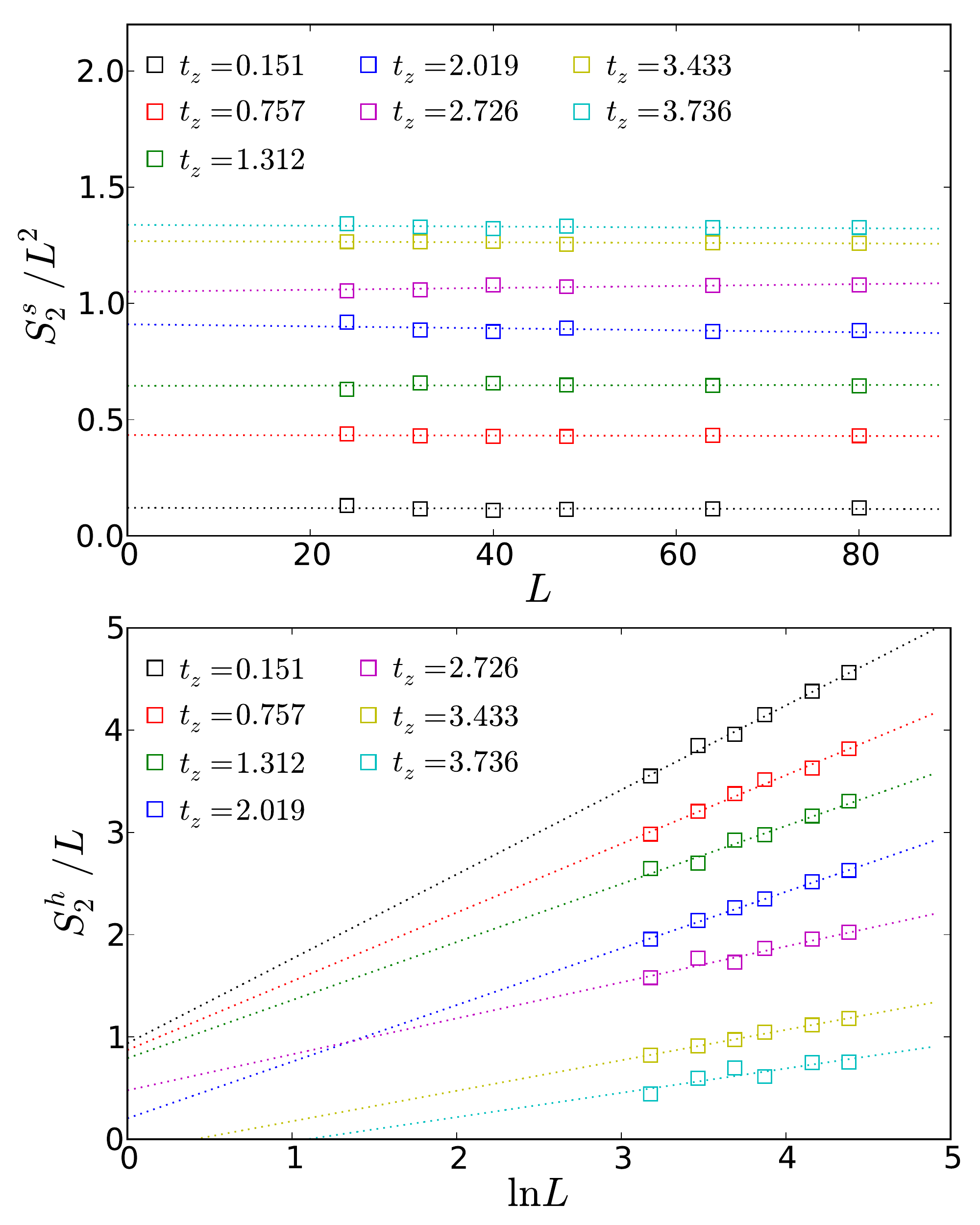}
\caption{(Color online)
Scaling properties of $S_2^s$ (top) and $S_2^h$ (bottom) for the non-interacting bilayer model.
Dotted lines represent linear fits to the data extracted from Fig.~\ref{fig:S2.vs.tz.free} at 
selected $t_z$'s.}
\label{fig:S2.vs.L.free}
\end{figure}

Next we move on to examine scaling properties of $S_2$.
In the band insulator phase, the previously mentioned result $S_2^s=2L^2\ln 2$ 
indicates that $S_2^s$ scales as the ``volume'' of ${\cal A}$, i.e. $L^2$. 
To examine the scaling behavior of $S_2^s$ in the metallic phase, we plot in the 
top panel of Fig.~\ref{fig:S2.vs.L.free} $S_2^s/L^2$ versus the linear dimension 
$L$ of the subsystem for $t_z < 4.0$. 
The data are fitted to a linear function and the results are shown as dotted lines 
in the figure. Within regression uncertainties, the slope of the fit is essentially zero, 
implying that $S_2^s\sim L^2$. 
In other words, $S_2^s$ obeys a ``volume law''. This result is a consequence of 
the fact that the interfacial area due to the partition scales as $L^2$.

The entanglement properties for the second partition (two half-layers) are
more intricate in the range $t_z < 4.0$. Since the orientation of the boundary 
between $\cal A$ and $\cal B$
and the normal vector of the Fermi surfaces are no longer perpendicular, the
behavior of $S_2^h$ is closely connected to the local geometry of Fermi surfaces.
In the bottom panel of Fig.~\ref{fig:S2.vs.L.free}, we show the scaling of $S_2^h$
in the metallic phase. Dotted lines in the figure are fits to the data according to
the formula $S_2^h/L = \alpha\ln L + \beta$. Instead of an ``area law'' where
$S_2^h\sim L$, the result of the fits shows that there is a logarithmic 
correction to the area law.
We note that in the bottom panel of Fig.~\ref{fig:S2.vs.L.free}, there are outliers
in the $S_2^h/L$ versus $\ln L$ plot at $t_z=1.312$, $2.726$, and $3.736$. These fluctuations
result from finite-size effects produced by ground state degeneracy at the Fermi level.
Such effects have strong dependence on $L$ and $t_z$.

\begin{figure}[h]
\includegraphics[scale=0.42]{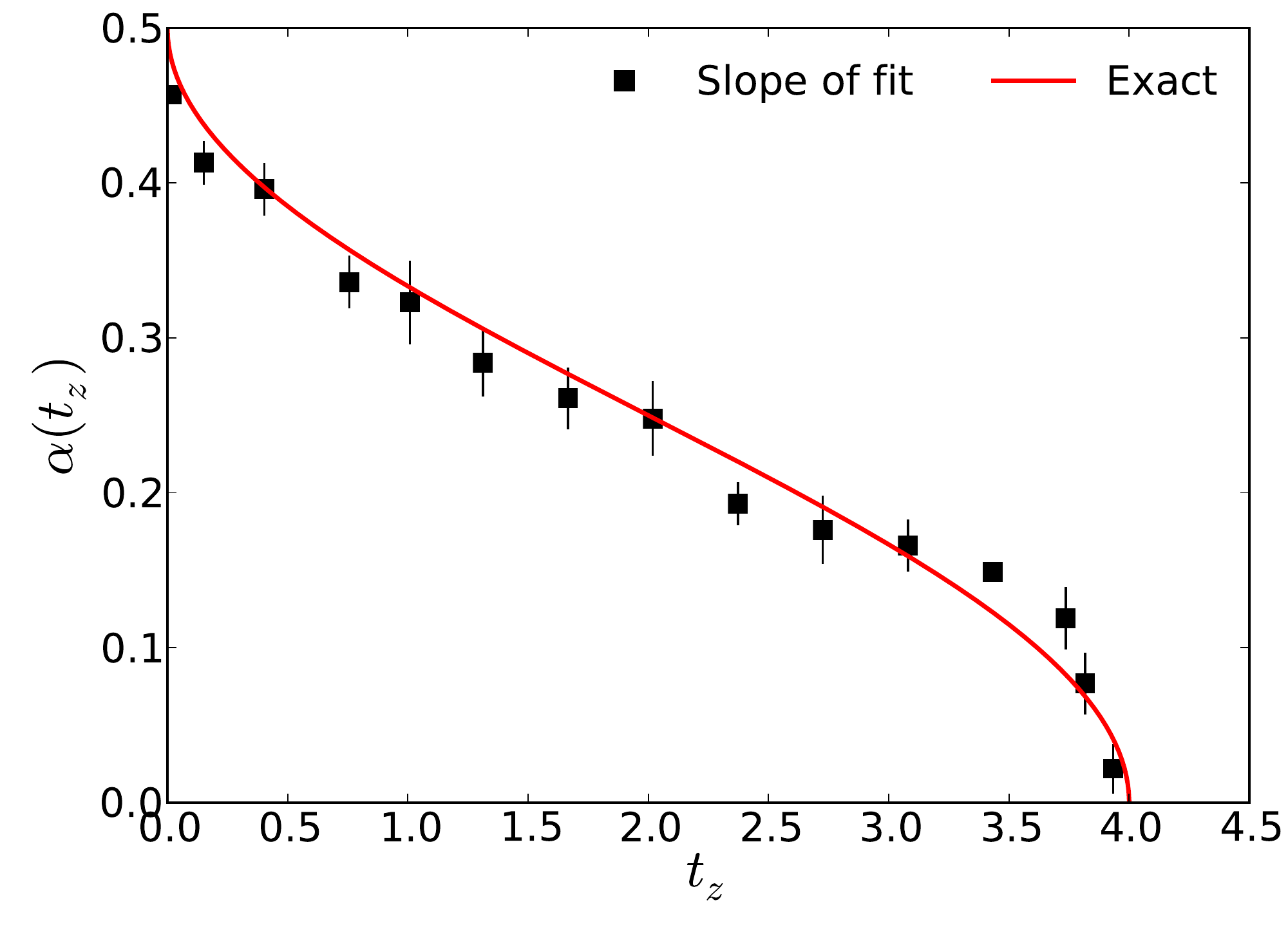}
\caption{(Color online)
Filled squares represent the coefficient of the fit $S_2^h/L=\alpha\ln L + \beta$ as a 
function of interlayer hopping $t_z$. The vertical line denotes the uncertainty of the 
regression. The solid curve is the function $\alpha(t_z)= 
1/(2\pi)\cos^{-1}(t_z/2-1)$ derived using Eq.~(6) for the non-interacting bilayer model.
}
\label{fig:Widom}
\end{figure}

For free fermions such a logarithmic correction to the area-law is expected.\cite{wolf}
The coefficient of the logarithmic correction has been derived in Ref.~\onlinecite{Gioev2006} 
based on the Widom conjecture:\cite{Widom1982}
\be
  \alpha(t_z) \sim \int_{\mbox{\scriptsize FS}}
              \left|\hat{n}_\bfx\cdot \hat{n}_\bfk\right| dS_\bfx dS_\bfk.
\ee
The integral is carried out over the surfaces of the subsystem and the Fermi surface. 
$\hat{n}_\bfx$ and $\hat{n}_\bfk$ are the unit normal vectors to the surfaces. 
In Fig.~\ref{fig:Widom}, we compare the exact result of free fermions and the 
coefficient $\alpha(t_z)$ extracted from the fits. The agreement is reasonably good
and supports the Widom conjecture for the bilayer tight-binding model.

\subsection{Bilayer Hubbard model}

\begin{figure*}[h]
\includegraphics[scale=0.44]{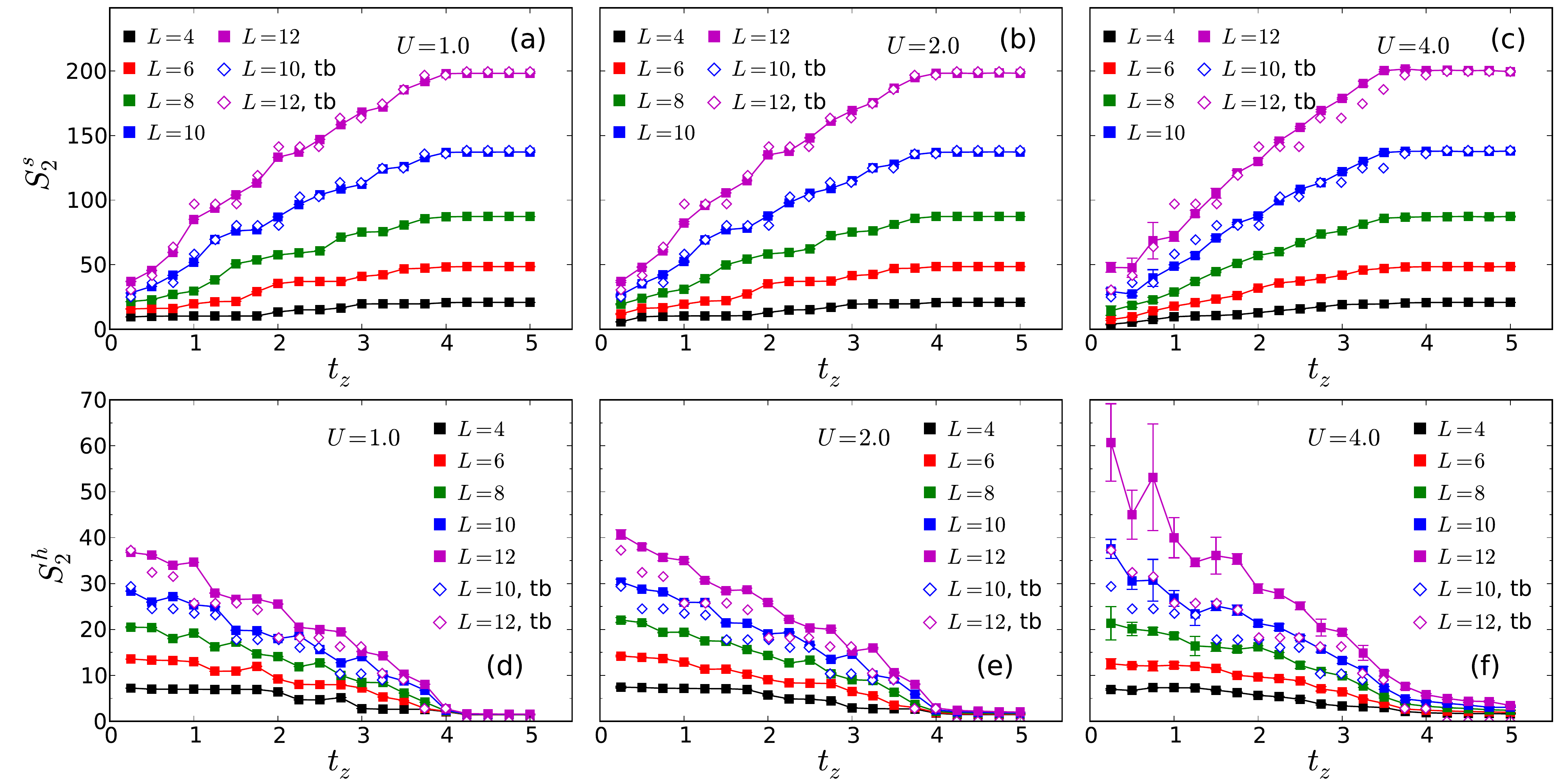}
\caption{(Color online) The second \Renyi entropy for the bilayer Hubbard model as a 
function of inter-layer hopping $t_z$. The upper and lower panels show 
$S_2^s$ and $S_2^h$ respectively. Simulations are performed
at temperature $T/t = 0.05$. Each data point is obtained by averaging 
over periodic and anti-periodic boundary conditions. In (a)--(c), the 
empty diamond points represent the ground state $S_2^s$ 
for the bilayer Hubbard model in the tight-binding (tb) limit $U=0$ 
computed using the correlation matrix technique.
}
\label{fig:S2.vs.tz}
\end{figure*}

\begin{figure*}[h]
\includegraphics[scale=0.44]{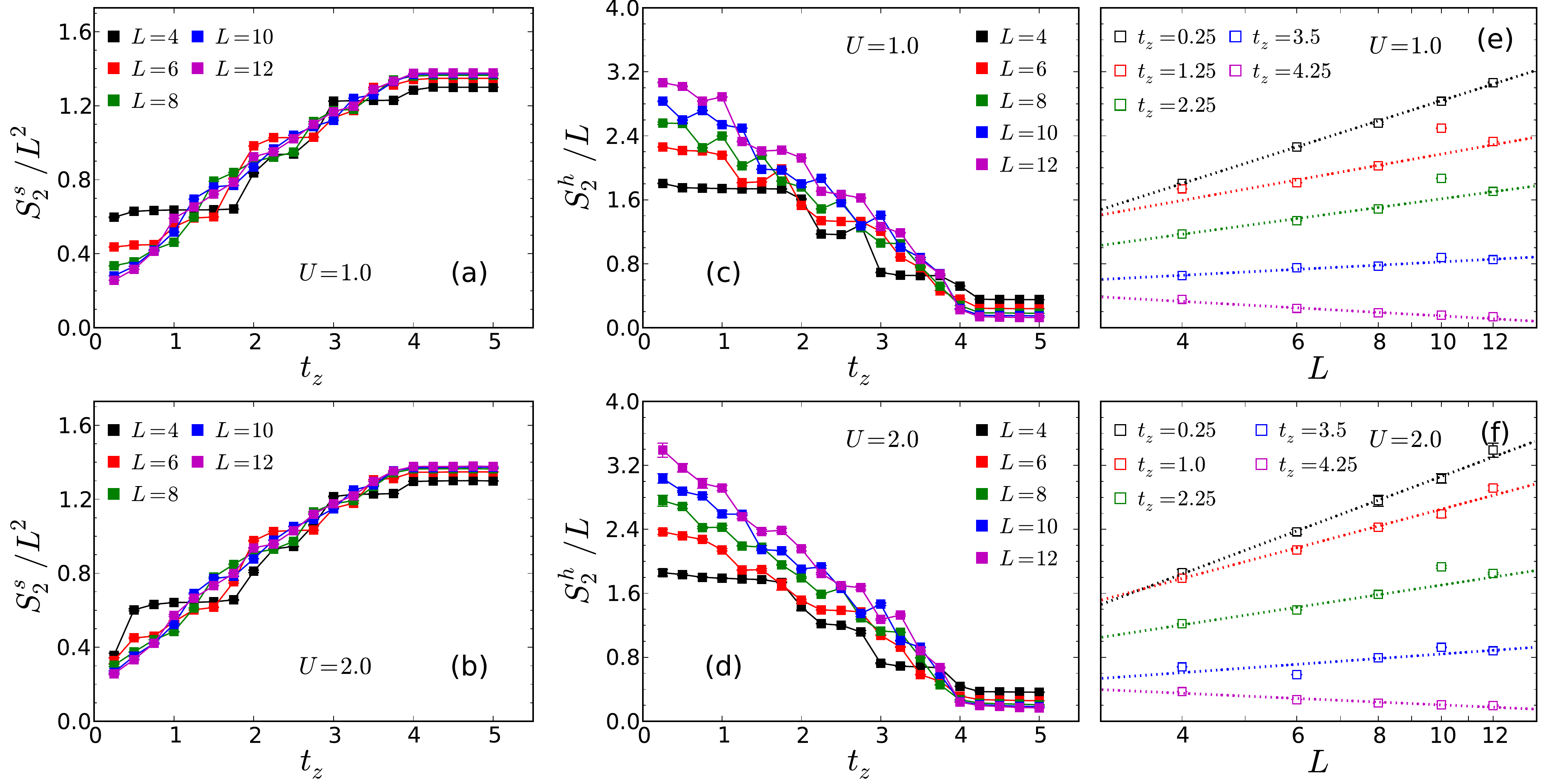}
\caption{(Color online) 
Scaling analysis of the \Renyi entropy for the bilayer Hubbard model at $U=1$ and 2.
Panel (a) and (b): $S_2^2/L^2$ versus $t_z$. Both figures show good data collapse,
suggesting a ``volume law'' for $S_2^s$. 
Figure (c) and (d): $S_2^h/L$ is plotted against $t_z$. If the system has an AF order,
$S_2^h$ is expected to have a linear scaling in $L$. The figures, however, do not support
the claim for small $t_z$. In (e) and (f), $S_2^h/L$ is fitted to an ansatz $\alpha\ln L + \beta$.
Dotted lines are results of the fit at several representative $t_z$ values. These regression
results indicate a possible logarithmic correction to the ``area law''.
}
\label{fig:S2.scaling}
\end{figure*}

The behavior of the second \Renyi entropy for the bilayer Hubbard model is 
shown in Fig.~\ref{fig:S2.vs.tz} at temperature $T/t=0.05$.
We have done simulations at different temperatures and made sure 
the results are not changing with temperature when $T/t$ reaches 0.05.
As indicated by Eq.~(\ref{eq:S2Grover}), the method relies on computing the
determinant of single particle Green's functions, which become ill-conditioned
at large $U/t$ values\cite{Bai2009}. Therefore the interaction strength will
be constrained in the range $U/t \leq 4.0$ in our work.

Using Eq.~(\ref{eq:S2Grover}), the Monte Carlo procedure accumulates statistics 
for the trace of the squared reduced density matrix 
$\Tr(\rho_{\cal A}^2)$, and the \Renyi entropy is a derived quantity from
the final results. To obtain a reliable estimation of $S_2$ and minimize 
possible bias, we use the jackknife resampling method to estimate $S_2$ and 
its statistical error. We also average the data over periodic and anti-periodic 
boundary conditions in order to reduce finite-size effects. 

Fig.~\ref{fig:S2.vs.tz} (a)-(c)  summarize $S_2$ for the single-layer 
partition at $U=1.0$, 2.0, and 4.0. The overall behavior of $S_2^s$ as a
function of $t_z$ is similar to that for free fermions: the entanglement entropy 
increases monotonically with $t_z$ and saturates when $t_z \gtrsim t_z^c$. 
For comparison, we also plot $S_2^s$ of free fermions for $L=10$ and 12. 
The data (represented by empty diamonds) is consistent with the bilayer 
Hubbard model results, albeit finite-size fluctuations are much stronger
in the free fermion data.
At a given $L$, the maximal value of $S_2^s$ for the bilayer Hubbard model
agrees with the free fermion result. This, again, is due to the fact
that $S_2^s$ only picks up short-range entanglement between singlet pairs
across the two layers. We speculate that the observed critical $t_z^c$ in 
the bilayer Hubbard model also corresponds to the place where the singlets 
start to break-up, indicating the onset of the singlet to AF transition.
For $U=1.0$ and 2.0, $t_z^c\sim 4.0$; while at $U=4.0$, the entanglement 
entropy plateaus at a slightly lower $t_z^c \sim 3.5$. Interestingly, these 
values are very close to the AF-BI transition phase boundary predicted by 
the VMC study.\cite{Ruger2014}

Next we turn our attention to the two half-layers partition. Simulations
are carried out at weak to moderate coupling strengths. Although details 
of the phase diagram in this parameter range are still under debate, it is 
generally agreed that at $U \gtrsim 4$, there is a direct AF to BI transition.
\cite{Bouadim2008,Kancharla2007,Ruger2014}
In the Heisenberg limit where $U\rightarrow\infty$, however, the phase 
transition is well characterized.\cite{heisenberg-qmc,Weihong1997,Hamer2012} 
Recently it has also been shown that the leading area-law coefficient shows 
a local maximum at the quantum critical point,\cite{Helmes2014} signaling the 
phase transition in the Heisenberg square-lattice bilayer.

The \Renyi entropy $S_2^h$ as a function of the inter-layer hopping $t_z$ is
plotted in Fig.~\ref{fig:S2.vs.tz} (d)-(e), at temperature $T/t=0.05$ and $U=1.0$, 
2.0, and 4.0 respectively. As a reference, the free fermion $S_2^h$ data is 
also shown in the same figures for $L=10$ and 12 (empty diamonds). 
The plateau-like structure in the free-fermion data is caused by the degeneracy at
the Fermi surfaces on a finite lattice. At $U=1$, the $S_2^h$ data still show kinks 
near the edge of the $U=0$ plateaus. As the interaction strength is increased,
the kinks become less pronounced. This suggests that the $U=0$ finite-size fluctuations 
can get carried over at small $U$s and produce the kinks.
A general trend of $S_2^h$ for the
system is that it reduces with increasing $t_z$. However, unlike the free 
fermion case, here the \Renyi entropy converges at a very slow rate to a low value
for $t_z \gtrsim 4.0$.
The comparison between the free fermion and the Hubbard model data suggests 
that $S_2^h$ gets enhanced, particularly in the region $t_z \lesssim 3.0$,
by increasing $U$. It is likely that the enhancement in the \Renyi entropy
is due to antiferromagnetic correlations developed across the two subsystems 
when the interaction is increased.

In Fig.~\ref{fig:S2.scaling} (a) and (b), $S_2^s$ is rescaled by $L^2$ and
plotted against $t_z$ for $U=1.0$ and 2.0 respectively. While the data have
finite-size fluctuations at small $L$'s, the figures show reasonably good
data collapse, indicating that $S_2^s$ scales as the ``volume'' of the subsystem,
just as in the free fermion case.
In the case of $S_2^h$, we plot in Fig.~\ref{fig:S2.scaling} (c) and (d)
$S_2^h/L$ as a function of inter-layer hopping at $U=1.0$ and 2.0. If the
system has AF ordering in this parameter range, as demonstrated by the VMC
results,\cite{Ruger2014} the entanglement entropy is expected to scale linearly
as $L$. In the region $0 < t_z < 4.0$, the figures seem to suggest that $S_2^h$
scales faster than $L$. In Fig.~\ref{fig:S2.scaling} (e) and (f), $S_2^h/L$
is fitted to a linear function $\alpha \ln L + \beta$ for several values of
$t_z$. The results for $t_z < 4.0$ suggest that a logarithmic correction to 
the linear scaling of $S_2^h$ might be possible.

However, in order to differentiate behavior of $S_2^h$ near the phase transition, 
it is necessary to carry out finite-size scaling study of the \Renyi entropy for 
larger system sizes. In particular, if the logarithmic correction to the linear
scaling of $S_2^h$ can be confirmed at $U\lesssim 2.0$, this would indicate possible 
existence of a Fermi surface. Likewise, a linear scaling of $S_2^h$ is expected
in the AF phase when $U$ is sufficiently large compared to the bandwidth.
As can be seen from Fig.~\ref{fig:S2.vs.tz}, our data suffers from strong statistical 
uncertainties in the region of interest. The 
fluctuations seem to grow with the lattice dimension $L$ and the coupling 
strength $U$, making it challenging to simulate large lattices at low temperatures.

We recall that the formalism presented by Eq.~(\ref{eq:S2Grover}) requires an 
explicit computation of the determinant of one-particle Green's functions. 
At zero temperature, it has been pointed out that the determinant may not 
exist\cite{Assaad2013} and a thermal broadening scheme is proposed to mend the 
issue.\cite{Assaad2013} At finite temperatures, it is known\cite{Bai2009} that, even at moderate 
coupling strengths, the Green's function matrix is ill-conditioned at low 
temperatures. As an example, the ratio of the largest and smallest eigenvalues 
of the inverse of the Green's function is of the order $10^{80}$ on a $16\times 16$ 
square lattice with $U=6.0$ and temperature $T/t = 0.067$.\cite{Bai2011} This
was the primary issue that plagues large scale DQMC simulations at low temperatures.
An efficient matrix decomposition scheme has been proposed to treat the instability 
issue inherited in DQMC and allows large scale simulations to be carried out
successfully.\cite{Bai2011,Varney2009} A similar stabilization scheme will be 
necessary in order to extract $S_2$ by directly computing determinants.

Very recently, an alternative method of measuring the entanglement entropies has been 
proposed.\cite{Broecker2014} The method expresses the \Renyi entropy as a logarithmic 
function of the ratio of two partition functions which can be sampled directly within 
DQMC scheme. This technique avoids the need of computing determinants and has been
shown to be more accurate than Eq.~(\ref{eq:S2Grover}).

\section{Summary}

In this work, we have studied entanglement properties across the AF-BI transition
in the bilayer Hubbard model. We focus on two bipartitions illustrated in
Fig.~\ref{fig:partitions}. Using the correlation matrix method, we have demonstrated
that in the tight-binding limit, the second \Renyi entropy show sharp signals
in the metal-BI transition. For the single-layer partition, the entanglement entropy
follows a strict ``volume law'' due to the singlet formation across the layers. For the
two half-layers partition, we were able to show a logarithmic break-down of the ``area law''
and confirm the prediction based on Widom's conjecture. 
For the bilayer Hubbard model, we have identified the value of $t_z$ that corresponds
to the break-up of singlet pairs across the bilayer, signaling the onset of AF to BI
phase transition. However, we were not able to pinpoint the critical point using the 
entanglement entropy data because of large statistical uncertainties and limited size 
of simulation cells. Finally, we have commented on the challenges of computing $S_2$ 
using the formalism adopted in our study.

\begin{acknowledgements}
CCC and RTS are supported by the DOE grant under the contract number DE-NA0001842-0 
and the University of California Office of the President.
RRPS acknowledges the support from NSF grant DMR-1306048.
\end{acknowledgements}

\bibliography{reference}

\end{document}

%% file: MyCommand.tex
\newcommand{\be}{\begin{equation}}
\newcommand{\ee}{\end{equation}}
\newcommand{\ba}{\begin{eqnarray}}
\newcommand{\ea}{\end{eqnarray}}

\newcommand{\upa}{\uparrow}
\newcommand{\dna}{\downarrow}

\newcommand{\COMMENTED}[1]{}

\newcommand{\Tr}{{\mbox{Tr}}}
\newcommand{\ket}[1]{|#1\rangle}
\newcommand{\bra}[1]{\langle #1|}

\newcommand{\ca}[2]{c_{#1 #2}^\dagger}
\newcommand{\de}[2]{c_{#1 #2}^{\phantom{\dagger}}}

\newcommand{\ob}[1]{{\langle #1\rangle}}

\newcommand{\bfi}{\mathbf{i}}
\newcommand{\bfj}{\mathbf{j}}
\newcommand{\bfk}{\mathbf{k}}

\newcommand{\bfx}{\mathbf{x}}

\newcommand{\bfI}{\mathbf{I}}